\documentclass[reprint,a4paper,showpacs,showkeys,pra,aps]{revtex4-1}
\usepackage[utf8]{inputenc}
\usepackage[english]{babel}
\usepackage{graphicx}
\usepackage{amsmath}
\usepackage{amssymb}
\usepackage{textcomp}
\usepackage{subfigure}
\newcommand{\Bipi}{$\mathrm{Bi}^{+}$}
\newcommand{\Biiip}{$\mathrm{Bi}_2^{+}$}
\newcommand{\Vac}{\mbox{$\mathrm{V_{\mathrm{Cl}}}$}}
\newcommand{\BiVac}{\mbox{$\mathrm{Bi}^{+}\cdots\mathrm{V_{\mathrm{Cl}}^{-}}$}}
\newcommand{\Cub}[1]{\mbox{$\mathrm{#1}\mathrm{m}\overline{3}\mathrm{m}$}}
\newcommand{\mkm}{\textmu{}m}
\begin{document}
\title{Near-infrared luminescence in bismuth-doped \mbox{TlCl}{} crystal}
\author{V.~G.~Plotnichenko}
\author{V.~O.~Sokolov}
\email[E-mail:~~]{vence.s@gmail.com}
\author{D.~V.~Philippovskiy}
\author{E.~M.~Dianov}
\affiliation{Fiber~Optics~Research~Center of the~Russian~Academy~of~Sciences \\
38~Vavilov~Street, Moscow 119333, Russia}
\author{I.~S.~Lisitsky}
\author{M.~S.~Kouznetsov}
\author{K.~S.~Zaramenskikh}
\affiliation{State Scientific-Research and Design Institute of Rare-Metal
Industry "Giredmet" JSC \\
5-1 B.Tolmachevsky lane, Moscow, 119017, Russia}

\begin{abstract} 
Experimental and theoretical studies of spectral properties of crystalline
\mbox{TlCl:Bi} are performed. Two broad near-infrared luminescence bands with a
lifetime about 0.25~ms are observed: a strong band near 1.18~\mkm{} excited by
0.40, 0.45, 0.70 and 0.80~\mkm{} radiation, and a weak band at $\gtrsim
1.5$~\mkm{} excited by 0.40 and 0.45~\mkm{} radiation. Computer modeling
of Bi-related centers in \mbox{TlCl}{} lattice suggests that \BiVac{} center
(\Bipi{} in Tl site and a negatively charged Cl vacancy in the nearest anion
site) is most likely responsible for the IR luminescence.
\end{abstract}
\maketitle

\section{Introduction}
For more than a decade bismuth-doped glasses and optical fibers attract a
considerable interest due to broadband IR luminescence in the range of
1.0--1.7~\mkm{} used in fiber lasers and amplifiers (see e.g. the review
\cite{Dianov09}). Although the origin of IR luminescence is still not clear,
recently the belief has been strengthened that subvalent Bi centers are
responsible for the luminescence \cite{Peng11}. In our opinion, monovalent
Bi centers are of a particular interest. Crystalline halides of monovalent
metals are convenient as model hosts to study such centers. These crystals have
a simple structure (primitive, \Cub{P}, or face-centered, \Cub{F}, cubic
lattice). Bismuth can easily form monovalent substitutional centers in such
lattice. Similar subvalent Tl and Pb centers in \Cub{F}{} crystals were studied
extensively (e.g. Tl in \mbox{KCl} \cite{Mollenauer83} and Pb in \mbox{MF$_2$}
(M$\, = \,$Ca, Sr, Ba) \cite{Fockele89}). By analogy, the models of Bi-related
centers in oxide glasses for fiber optics were suggested \cite{Dianov10}.
Subvalent Bi centers in cubic halide crystals were studied for the first time in
\mbox{BaF$_2$} \cite{Su09} and then in \mbox{CsI} (\Cub{P}) \cite{Su11, Su12}.
\begin{figure}
\includegraphics[scale=0.50, bb=55 280 550 770]{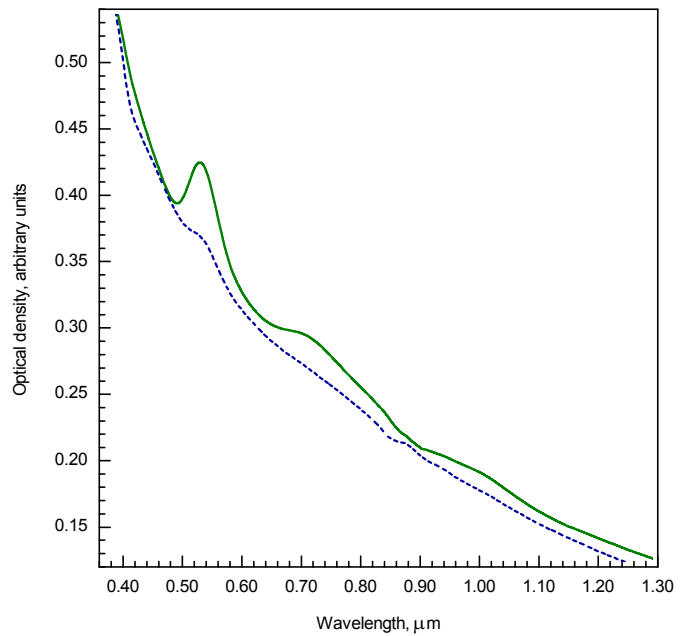}
\caption{%
Optical density of \mbox{TlCl:Bi}{} crystals with a higher (solid line) and
lower (dashed line) Bi content
}
\label{fig:figure1}
\end{figure}

In the present work we report the results of spectroscopic study and computer
modeling of Bi-related centers in bismuth-doped \mbox{TlCl}{} crystal (\Cub{P}).
\begin{figure*}
\includegraphics[scale=1.045, bb=63 270 549 768]{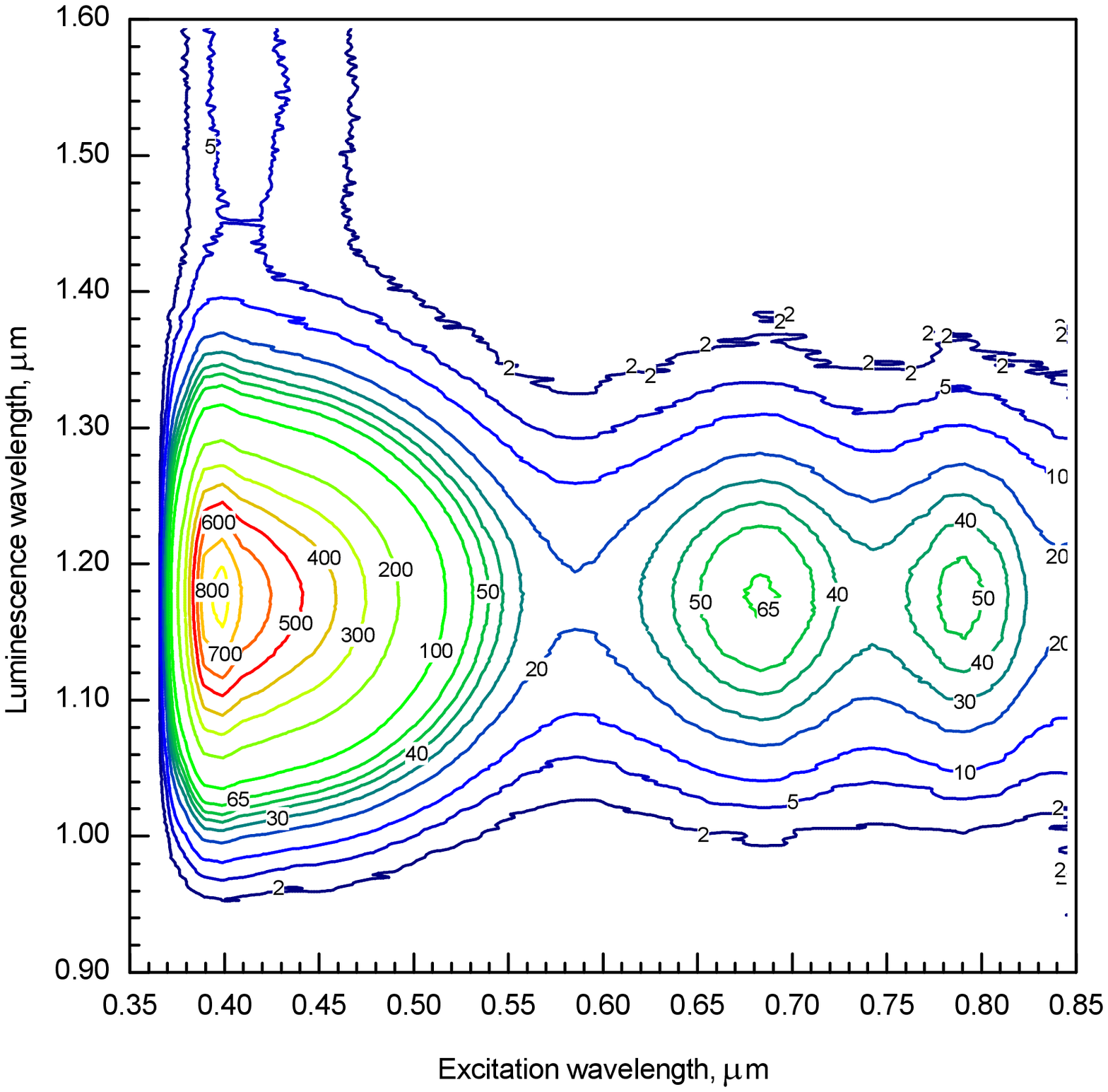}
\caption{%
Bi-related luminescence in \mbox{TlCl:Bi}{} crystal (intensity is given in
arbitrary units)
}
\label{fig:figure2}
\end{figure*}

\section{Experimental}
\mbox{TlCl}{} single crystals were grown by Bridgman-Stockbarger method as
described in \cite{Lisitsky04}. \mbox{TlCl}{} with a cationic impurities
content of $\lesssim 1 \times 10^{-4}$~wt.\% was used as an initial material.
\mbox{$\mathrm{BiCl}_3$}{} (99.999~\%, ultra dry) with concentration of
0.1--0.4~wt.\%  was used as dopant. Crystals were grown in nitrogen atmosphere
(residual pressure of $10^{-2}$~mm~Hg) in Pyrex glass ampules with a rate of
$\lesssim 2$~mm/h. The interface--phase boundary was kept at 15--20~mm above the
separating diaphragm. The temperatures of the upper and lower zones of a
vertical furnace were held at $460^\circ$C ($30^\circ$C above \mbox{TlCl}{}
melting temperature) and at $350^\circ$C, respectively. Crystals of 22~mm
diameter and 80--100~mm length were produced. For spectroscopic measurements
5~mm-thick plates were cut from the crystals, ground and polished. The
measurements were carried out by using two samples with different total Bi
content. The transmission spectra were measured by Perkin Elmer Lambda~900
spectrophotometer. The luminescence emission and excitation spectra, as well as
the luminescence lifetimes were measured by Edinburgh Photonics FLS~980
spectrometer.

One strong band near 0.53~\mkm{} and three weak bands near 0.45, 0.72, and
1.0~\mkm{} are observed in the absorption spectrum (Fig.~\ref{fig:figure1}). Two
IR bands are observed in the luminescence spectrum (Fig.~\ref{fig:figure2}). The
strong band near 1.18~\mkm{} with a half-width about 0.2~\mkm{} is excited in
0.40, 0.45, 0.70, and 0.80~\mkm{} absorption bands. The weak band at $\gtrsim
1.5$~\mkm{} is excited in 0.40 and 0.45~\mkm{} absorption bands. The
luminescence lifetime is 0.20--0.35~ms in both IR bands. It should be emphasized
that the luminescence excitation spectrum differs from the absorption spectrum,
and that the absorption increases with Bi content (Fig.~\ref{fig:figure1}), as
opposed to the IR luminescence.
\begin{figure*}
\subfigure[]{%
\includegraphics[scale=0.90, bb=270 80 452 680]{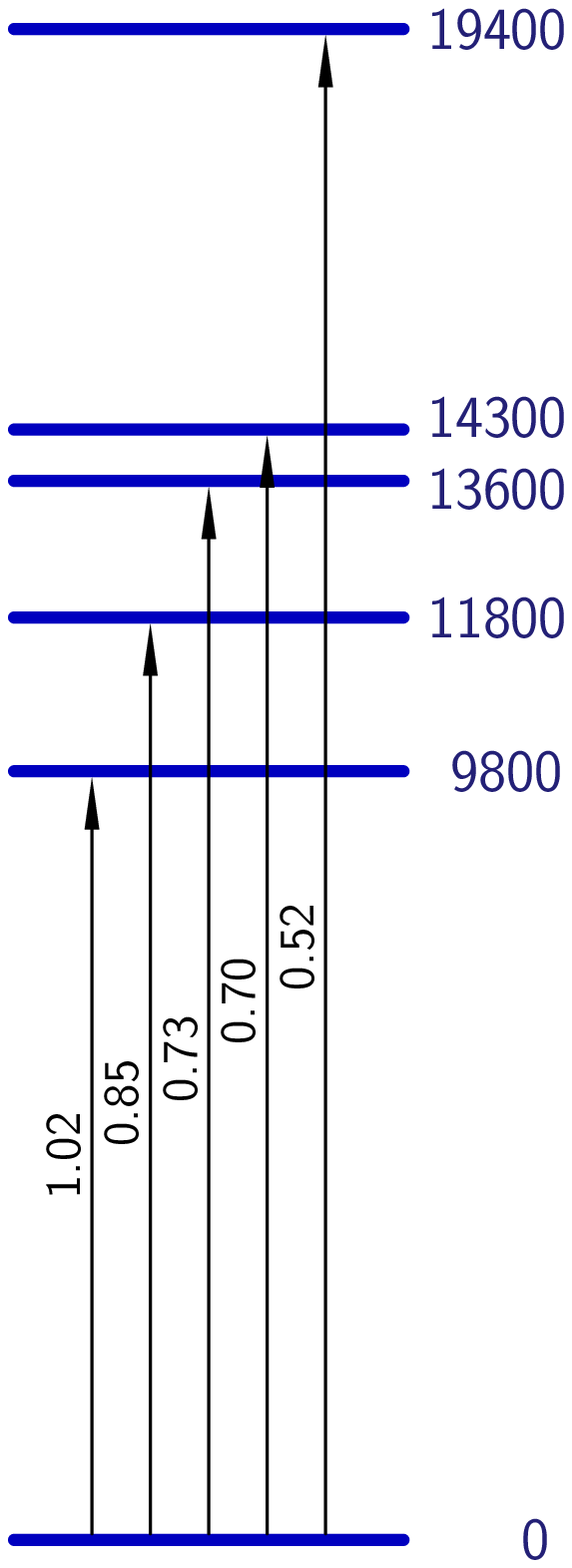}
\label{fig:figure3a}
}
\subfigure[]{%
\includegraphics[scale=0.90, bb=270 80 452 680]{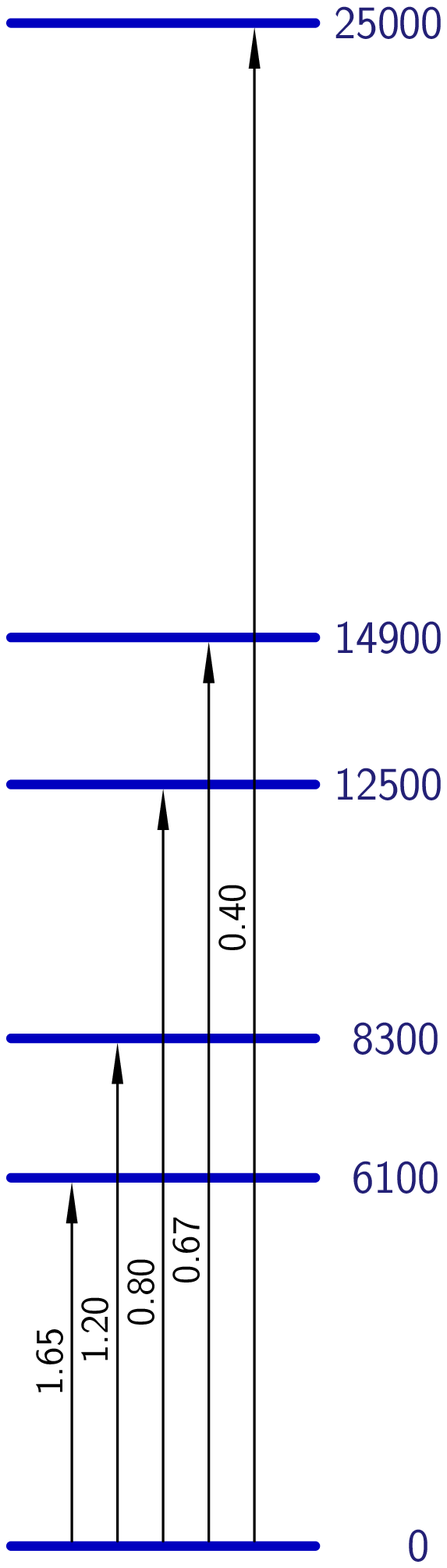}
\label{fig:figure3b}
}
\subfigure[]{%
\includegraphics[scale=0.90, bb=270 80 452 680]{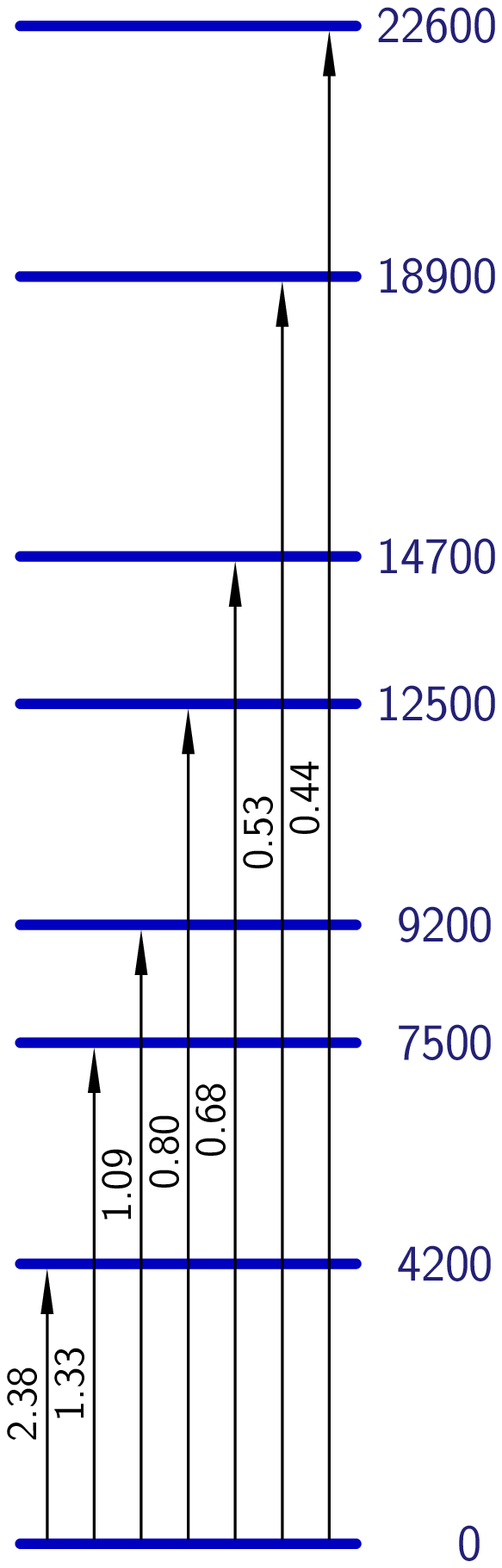}
\label{fig:figure3c}
}
\caption{%
Calculated levels and transitions of Bi-related centers in \mbox{TlCl:Bi}{}
crystal:
\subref{fig:figure3a}~\Bipi{} substitutional center,
\subref{fig:figure3b}~\BiVac{} complex,
\subref{fig:figure3c}~\Biiip{} dimer center (level energies are given in
$\textrm{cm}^{-1}$, transition wavelengths in \mkm)
}
\label{fig:Bi_centers_levels}
\end{figure*}

\section{Modeling of bismuth-related centers in TlCl:Bi}
To understand the origin of the IR luminescence, we performed a computer
simulation of the structure and absorption spectra of several Bi-related centers
possibly occurring in \mbox{TlCl:Bi}{} crystal. Firstly, monovalent Bi
substitutional center, \Bipi, was studied as the main form of Bi in \mbox{TlCl}.
Further, \BiVac{} complex formed by the \Bipi{} center and negatively charged Cl
vacancy in its first coordination shell (in other words, both located in the
neighboring sites of \mbox{TlCl}{} lattice) was modeled as a center similar to
\mbox{$\textrm{Tl}^0\!\left(1\right)$} center in \mbox{KCl}{} crystal
\cite{Mollenauer83}. And finally, dimer center, \Biiip, formed by two \Bipi{}
sustitutional centers in the nearest sites of \mbox{TlCl}{} cation sublattice
with an extra electron, was chosen by analogy with the assumptions made in
\cite{Su11, Su12} for \mbox{CsI:Bi}.

The modeling was performed in supercell approach. To model \Bipi{} and \BiVac{}
centers, $3 \times 3 \times 3$ \mbox{TlCl}{} supercell containing 54 atoms was
chosen, and $3 \times 3 \times 4$ supercell with 72 atoms was used for \Biiip.
In the central region of the supercell certain Tl atoms were substituted by Bi
atoms and an anion vacancy was formed by a removal of one Cl atom. Charged
centers were simulated changing the total number of electrons in the supercell.
Equilibrium configurations of the centers were found by a complete optimization
of the supercell parameters and atomic positions with the gradient method. All
calculations were performed using \mbox{Quantum-Espresso}{} package \cite{QE} in
the plane wave basis in the generalized gradient approximation of density
functional theory (DFT) with ultra-soft pseudopotentials built with PBE
functional \cite{PBE}. To test the approach, we calculated \mbox{TlCl}{} lattice
parameters both for \mbox{TlCl}{} unit cell and for the supercells with both
atomic positions and cell parameters completely optimized. The results
convergence was tested with respect to the plane wave cutoff energy and to the
$k$ points grid.

Configuration of the Bi-related centers calculated by this means was used to
calculate the absorption spectra of the centers by the Bethe-Salpeter equation
method based on all-electron full-potential linearized augmented-plane wave
approach. The calculation was performed using \mbox{Elk}{} code \cite{Elk} in
the DFT local spin density approximation with PW-CA functional \cite{LSDA-PW,
LSDA-CA}. Spin-orbit interaction essential for bismuth-containing systems was
taken into account. Scissor correction was applied in transition energies
calculation with the scissor value found using modified Becke-Johnson
exchange-correlation potential \cite{Becke06, Tran09, Tran11}. Convergence of
the results was tested with respect to plane wave cutoff energy, to the angular
momentum cutoff for the muffin-tin density and potential, and to the $k$ points
grid choice.

Configurational coordinate diagrams of the Bi-related centers were calculated in
a simple model restricted to the lowest excited states with a displacement of Bi
atom(s) along $\left[111\right]$ axis for \Bipi{} and \BiVac{} centers and along
$\left[001\right]$ axis for \Biiip{} center. In spite of the fact that the model
is inherently approximate, it shows that in all three centers studied the Stokes
shift corresponding to a transition from the first excited state to the ground
one do not exceed the accuracy of the excited state energy calculation. Hence it
is reasonable enough to estimate the IR luminescence wavelengths by taking the
Stokes shift to be zero.

Calculation of the \Bipi{} substitutional center shows that the crystal lattice
is distorted rather slightly: Bi atom lies in the cation site, the nearest Cl
atoms are displaced towards Bi atom, and the nearest Tl atoms are displaced
apart from Bi atom, so that Bi$\relbar$Cl and Bi$\relbar$Tl distances are
3.120 and 3.889~\AA{}, respectively (3.320 and 3.834~\AA{} in \mbox{TlCl}{}
crystal). A luminescence band is expected to be near 1.0~\mkm{} excited in
absorption near 0.8, 0.7, and $\sim 0.5$~\mkm{} (Fig.~\ref{fig:figure3a}).
Another one, with much a lower lifetime, may occur near 0.8~\mkm.

Lattice relaxation turns out to be much more significant in the \BiVac{} complex
center. Bi atom is displaced by 1.146~\AA{} from the cation site towards the
vacant Cl site, the nearest Cl and Tl atoms are displaced towards Bi atom, and
the Tl atoms surrounding Cl vacancy are displaced apart from the vacant site. So
the Bi$\relbar$Cl, Bi$\relbar$Tl, and Tl$\relbar$\Vac{} distances are 3.024,
3.598, and 4.009~\AA, respectively, as compared to 3.320, 3.834, and 3.320~\AA{}
in \mbox{TlCl}{} crystal. The relaxation is accompanied by the electron density
shifted from Bi atom into the Cl vacancy region, so that the complex center may
be thought of as a bound pair of ions, ``\Bipi{} plus negatively charged
V$_{\mathrm{Cl}}^{-}$ vacancy''. The IR luminescence is expected in the bands
near 1.6 and 1.2~\mkm, both excited in absorption near 0.8, 0.7, and $\sim
0.4$~\mkm{} (Fig.~\ref{fig:figure3b}). Again a luminescence band with
significantly shorter (by an order of magnitude) lifetime may occur near
0.8~\mkm.

As well our modeling shows that \Biiip{} dimer centers can occur in
\mbox{TlCl:Bi}{} crystal. Bi atoms are displaced from the adjacent cation sites
towards each other, the nearest Cl atoms are displaced towards the dimer,
and the nearest Tl atoms are displaced apart from the dimer. The Bi$\relbar$Bi,
Bi$\relbar$Cl, and Bi$\relbar$Tl distances are 2.903, 3.060, and 3.960~\AA{},
respectively (3.834, 3.320, and 3.834~\AA{}, respectively, in \mbox{TlCl}{}
crystal). The excess charge, $-1\left|\mathrm{e}\right|$, turns out to be
localized mainly in the Bi atom and marginally in the nearest Cl atoms. So the
center is \Biiip{} dimer indeed. One might expect the luminescence bands near
1.1 and 1.3~\mkm{} excited in absorption near 1.1, 0.8, 0.7, 0.5 and $\sim
0.4$~\mkm. In one more band at $\sim 2.4$~\mkm{} the luminescence corresponding
to the transition from the lowest excited state (Fig.~\ref{fig:figure3c}) might
be observable at a low temperature.

\section{Conclusions}
Our spectroscopic data and results of modeling of Bi-related centers in
\mbox{TlCl:Bi}{} crystal suggest that the near-infrared luminescence in
\mbox{TlCl:Bi}{} is caused mainly by \BiVac{} complexes formed by Bi
substitutional ions and intrinsic defects, chlorine vacancies
(Fig.~\ref{fig:figure2} and \ref{fig:figure3b}). The Bi-related contribution to
the total absorption is caused mainly by single \Bipi{} substitutional centers
not responsible for the IR luminescence (Fig.~\ref{fig:figure1} and
\ref{fig:figure3a}). So the IR luminescence excitation spectrum
(Fig.~\ref{fig:figure2}) differs significantly from the absorption spectra of
\mbox{TlCl:Bi}{} samples (Fig.~\ref{fig:figure1}). \Biiip{} dimer complexes in
\mbox{TlCl:Bi}{} can contribute perceptibly neither to the IR luminescence
spectra nor to the absorption spectra, as distinct from the assumptions on
\mbox{CsI:Bi}{} crystals \cite{Su11, Su12}.

\acknowledgments
The authors are grateful to S.~V.~Firstov for assistance in luminescence
measurements. This work was supported in part by Basic Research Program of the
Presidium of the Russian Academy of Sciences and by Russian Foundation for Basic
Research (grant 12-02-00907).
%
%

\end{document}